\documentclass[journal=jpclcd,manuscript=letter,layout=twocolumn]{achemso}
\usepackage{dcolumn}
\usepackage{amssymb,amsmath}
\usepackage{graphicx}
\usepackage{hhline}
\usepackage{textcomp}
\usepackage{ADDmacros}

\author{Vikram Plomp}
\altaffiliation{Contributed equally to this work}
\author{Xu-Dong Wang}
\altaffiliation{Contributed equally to this work}
\affiliation{Radboud University, Institute for Molecules and Materials, Heijendaalseweg 135, 6525 AJ 
Nijmegen, the Netherlands}
\author{Fran{\c c}ois Lique}
\affiliation{Universit{\'e} de Rennes, Institut de Physique de Rennes, 263 avenue du G{\'e}n{\'e}ral 
Leclerc, 35042 Rennes CEDEX, France}
\email{francois.lique@univ-rennes1.fr}
\author{Jacek K{\l}os}
\affiliation{University of Maryland, Department of Physics, Joint Quantum Institute, 
College Park, MD, United States of America}
\author{Jolijn Onvlee}
\author{Sebastiaan Y.T. van de Meerakker}
\email{basvdm@science.ru.nl}
\affiliation{Radboud University, Institute for Molecules and Materials, Heijendaalseweg 135, 6525 AJ 
Nijmegen, the Netherlands}

\title{High-resolution imaging of C + He 
collisions using Zeeman deceleration and VUV detection}

\begin{tocentry}
	{\includegraphics[width=5cm]{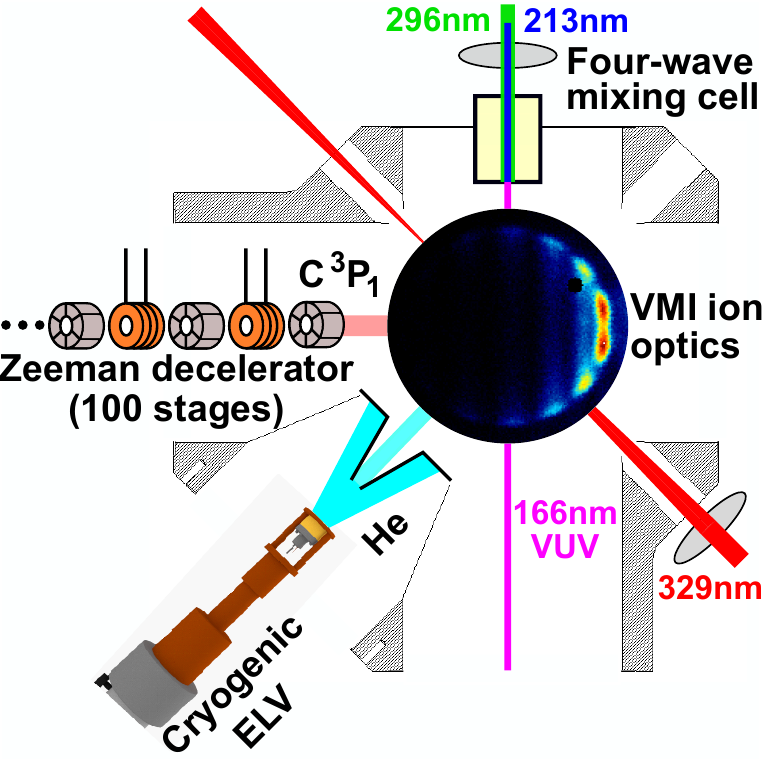}}
	\\\\
	The used combination of Zeeman deceleration, (1+1') (VUV+UV) near-threshold REMPI detection and velocity map imaging (VMI) allowed for high-resolution imaging of C-atom scattering distributions after interaction with He.
\end{tocentry}

\begin{document}

\begin{abstract}
	High-resolution measurements of angular scattering distributions provide a sensitive test for 
	theoretical descriptions of collision processes. Crossed beam 
	experiments employing a decelerator and velocity map imaging have proven 
	successful to probe collision cross sections with extraordinary 
	resolution.   
	However, a prerequisite to exploit these possibilities is the availability 
	of a near-threshold state-selective ionization scheme to detect the 
	collision products, which for many species is either absent or inefficient.
	We present the first implementation of recoil-free vacuum ultraviolet (VUV) based detection in 
	scattering experiments involving a decelerator and velocity map imaging. This allowed for 
	high-resolution measurements of state-resolved angular scattering distributions for 
	inelastic collisions between Zeeman-decelerated carbon C($^3P_1$) atoms and helium 
	atoms. We 
	fully resolved diffraction oscillations in the angular distributions, which showed excellent 
	agreement with the distributions predicted by quantum scattering 
	calculations. Our approach offers exciting prospects to investigate a large range of scattering processes with 
	unprecedented precision.
\end{abstract}

Acquiring a detailed understanding of molecular interactions 
is an 
important goal in physical 
chemistry \cite{LevineBernstein1987}. Theoretical descriptions for these interactions have become ever more 
advanced, and sophisticated experiments have been designed to test their quality 
\cite{Yang:ModTrendChemReactDyn, TutorialsMRD, Casavecchia2000, Kohguchi2002, Picard2019}.
In the last decades, crossed beam experiments that studied molecular collisions in the gas phase have 
provided sensitive probes for the potential energy surfaces (PESs) underlying 
molecular interactions \cite{Liu2006, Yang2011, Naulin2014, Aoiz2015}. 
One of the most stringent tests for the involved PESs can be found in 
measurements of 
angular scattering distributions, which directly reflect the differential cross sections 
(DCSs) \cite{Onvlee2015, Amarasinghe2020, Yuan2020, Chen2021, Paliwal2021}. 
In this regard, the high resolution afforded by the combination of Stark 
\cite{Meerakker2012,Meijer2021} or Zeeman 
\cite{Meerakker2012,Jansen2020,Heazlewood2021} deceleration to control collision partners and velocity 
map imaging (VMI) to probe collision products has enabled the observation of delicate features in 
angular scattering distributions
\cite{Onvlee2016,gao2018observation,Jongh2020,Plomp2020}. 
However, the sparse availability of efficient near-threshold resonance-enhanced multiphoton ionization 
(REMPI) schemes, a prerequisite for obtaining high-resolution scattering images, still limits the 
number of systems for which the full potential of this approach can be exploited. 

Of particular interest are scattering systems involving multiple interaction potentials with 
non-adiabatic 
couplings between them where the Born-Oppenheimer approximation no longer holds. A typical example is 
the 
spin-orbit 
(de-)excitation of ground-state atomic carbon, C($^3P_j$) $\rightarrow$ C($^3P_{j'}$), in collisions 
with He or H$_2$ 
\cite{Bergeat2018,Picard2002}.
This process is highly relevant for interstellar cloud cooling and plays an important role in 
chemical modelling of the interstellar medium \cite{Bensch2003,Neufeld1995}. To date, several experimental and theoretical studies concerning the collision-induced 
spin-orbit transitions in C($^3P_j$) atoms have been published
\cite{Bergeat2018,Bergeat2019,Klos2018,Picard2002,Monteiro1987,Lavendy1991,Staemmler1991}. In 
particular, Bergeat \etal reported experimental and theoretical 
integral cross 
sections (ICSs) for C($^3P_0$) + He $\rightarrow$ C($^3P_1$/$^3P_2$) + He collisions
\cite{Bergeat2018}. The excellent agreement between experiment and theory allowed a detailed 
description of resonance features in the collision energy dependent ICSs. 
Although currently unavailable, high-resolution experimental investigations of quantum-state-resolved 
DCSs for these processes
could provide even 
more stringent tests for theory as well as 
further insight into the 
underlying scattering mechanisms. The main 
bottleneck here is the 38 m/s ion recoil intrinsic to the conventional (2+1) UV REMPI detection of 
C($^3P_j$) 
\cite{Bergstroem1989,Jankunas2015}, which would wash out the fine details that can be observed in the
experimental scattering distributions. 

In this work, we experimentally probed state-resolved DCSs for the spin-orbit de-excitation 
collision process C($^3P_1$) + He $\rightarrow$ C($^3P_0$) + He with high resolution by the first-time implementation of VUV-based REMPI
detection in a crossed beam experiment employing a decelerator and VMI. The C atom is well suited for manipulation using magnetic fields \cite{Jankunas2015, Karpov2020}, and thus we used a Zeeman 
decelerator to prepare velocity-controlled packets of C($^3P_1$) atoms with narrow velocity 
and angular spreads. Despite the challenges arising from the reduced carbon beam density 
after deceleration and the low VUV power generated by difference frequency mixing, scattered carbon 
atoms were efficiently detected without ion-recoil by implementing a (1+1') (VUV+UV) REMPI scheme. 
The resulting exceptional resolution allowed us to fully resolve diffraction oscillations, for which 
excellent agreement was found with simulations based on \textit{ab initio} calculations of the 
involved potential energy curves (PECs). Since the use of VUV light for REMPI detection is generally 
applicable \cite{Kung1991, 
Ng2014} and provides the perspective of recoil-free detection for many atomic and molecular species, 
this approach
offers exciting prospects to study a large range of collision processes with an unprecedented 
level of precision.

The recoil-free (1+1') (VUV+UV) REMPI scheme for the state-selective detection of
C($^3P_j$) atoms is 
schematically depicted in \autoref[(a)]{fig:REMPI}. It employs the 2p3s $^3P_j$ $\leftarrow$ 
2p$^2$ $^3P_j$ transition induced by 166 nm VUV light, which is produced by difference frequency mixing ($2\omega_1 - \omega_2$) 
\cite{Hilber1987,Marangos1990} of co-propagating $\lambda_1 = 212.56$~nm 
($0.5$ mJ) and $\lambda_2 = 296$~nm ($0.4$ mJ) laser beams focussed inside a gas cell 
filled with 85 mbar krypton. While this transition is well known from precision spectroscopy experiments \cite{Glab1998,Lai2020}, it has not been used for REMPI detection in previous scattering experiments. The ion-recoil associated with the excess energy of ionization by e.g. a $\lambda_1$ or $\lambda_2$ photon is generally of no particular importance for spectroscopic investigations. 
When imaging scattering distributions, however, this ion-recoil induces a velocity blurring that washes out the 
fine structures that can be observed. Therefore, our implementation, which is similar to that of Glab 
\etal \cite{Glab1998}, uses $\lambda_3 \sim329$ nm light (8.5 mJ, partially focussed) for 
subsequent near-threshold ionization that allows for efficient high-resolution imaging of the carbon 
atoms. 

For the VUV excitation step, the 2p3s $^3P_1$ intermediate state was chosen to
state-selectively detect either the C($^3P_1$) from the decelerated beam or the 
C($^3P_0$) scattering product. The UV laser pulses (212, 296 and 329 nm) were generated by frequency 
doubling or tripling the output of three separate dye-lasers. Optimal time overlap for the 212 and 296 
nm laser pulses used to generate VUV light was ensured with the use of a shared Nd:YAG pump laser and 
suitable delay line. For the 329 nm ionization laser a separate Nd:YAG pump laser was used. The time 
jitter for each pump laser amounts to less than 1 ns, which is significantly shorter than both the 
laser pulse duration of around 6 ns full width at half maximum (FWHM) and the expected 2.7 ns lifetime of the 2p3s $^3 P_1$ 
intermediate state \cite{Haar1991, Zheng2002}. The VUV light was not separated from its 
parent $\lambda_1$ and $\lambda_2$ UV beams, and was focussed by a MgF$_2$ lens (focal 
length $\sim$ 275~mm) at a substantial distance behind the detection region.

\begin{figure}[!htb]
   \centering
    \resizebox{1.0\linewidth}{!}
    {\includegraphics{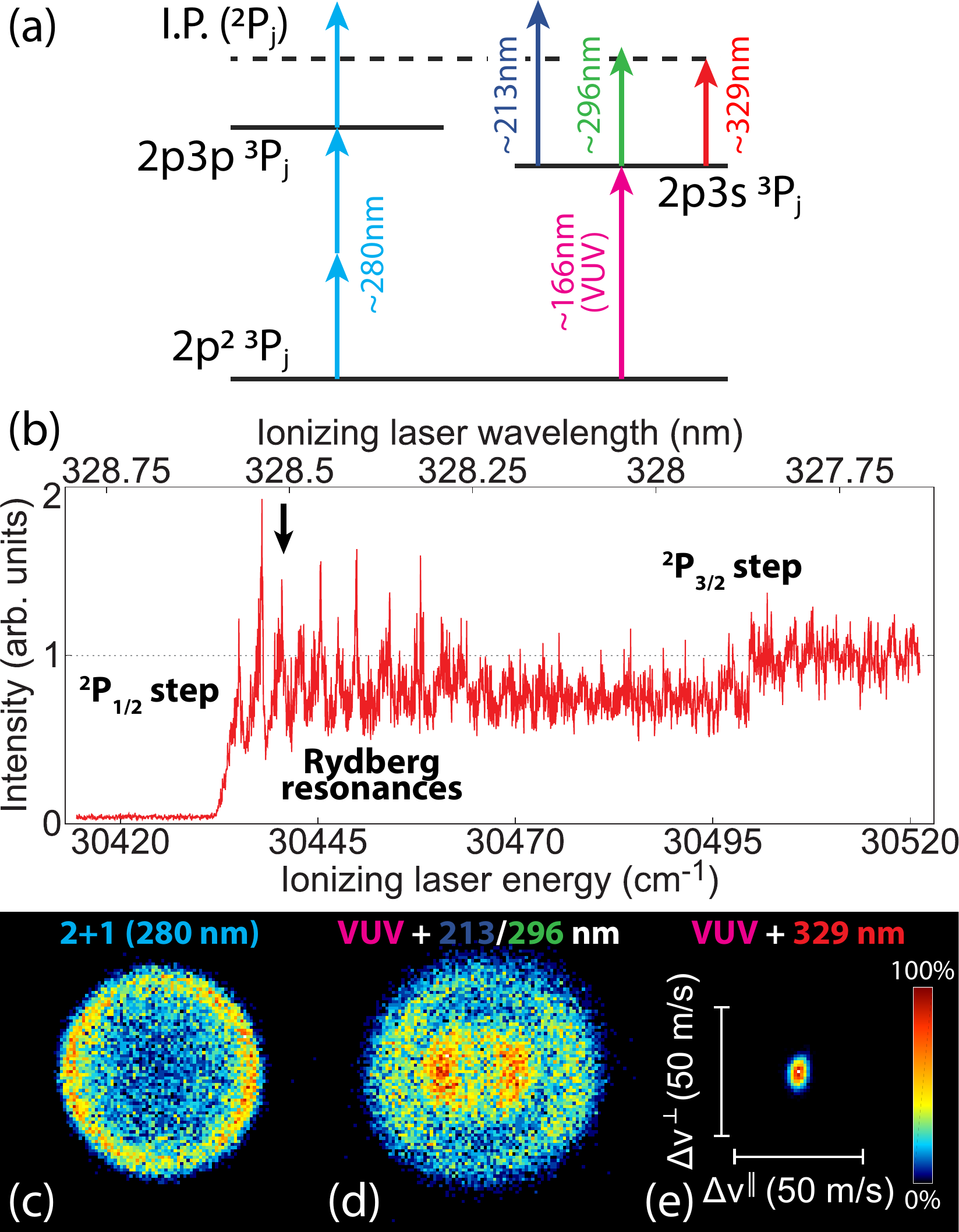}}
    \caption{(a) Schematic depiction of the conventional (2+1) REMPI scheme and different competing 
    ionization pathways for the (1+1') REMPI scheme involving VUV excitation. (b) The observed (1+1') REMPI yield as a 
    function of the additional $\lambda_3$ ionizing laser wavelength after 
    VUV excitation to the 2p3s $^3P_1$ state. The packet of 
    C($^3P_1$) atoms exiting the Zeeman decelerator with a mean longitudinal velocity of $\text{v}^\parallel=300$ m/s was imaged with 
    the conventional (2+1) REMPI scheme (c), and the (1+1') REMPI scheme both without (d) and with 
    (e) $\lambda_{3,\text{res}}$ radiation present. Each image pixel corresponds to a 
    velocity of around 1.2 m/s.}
    \label{fig:REMPI}
\end{figure}

When scanning the $\lambda_3$ ionization laser wavelength after VUV excitation, see 
\autoref[(b)]{fig:REMPI}, a clear step in the ion yield was observed for the threshold of both the $^2P_{1/2}$ ionic ground 
state as well as the $^2P_{3/2}$ state of the ion that lies just 63.42 cm$^{-1}$ higher in energy 
\cite{Moore1993}. Furthermore, just above the $^2P_{1/2}$ threshold a series of sharp peaks can be 
observed that correspond to resonant excitation to autoionizing Rydberg states \cite{Glab1998}. While these 
resonances 
provide a strong enhancement in signal level just above the ionization threshold, many of them lead to a 
small increase in blurring of the VMI images. This blurring is attributed to the expected long 
lifetimes 
of these Rydberg states in combination with the employed large ionization volume (several mm in each 
dimension), which poses especially demanding conditions for accurate velocity mapping.  However, the resonance at $\lambda_{3,\text{res}} = 328.5079$~nm, indicated by the arrow in 
\autoref[(b)]{fig:REMPI}, was found to give a significant increase in ion yield while 
causing only a marginal increase in image blurring. At this peak, the ratio of the signal levels with 
the $\lambda_{3,\text{res}}$ laser on and off was found to be around 33:1, which gives a lower bound 
for the ratio 
between signal from low-recoil near-threshold ionization and the high-recoil contribution from 
ionization by the $\lambda_1$ and $\lambda_2$ laser beams co-propagating with the VUV light. It should 
be noted 
that 
separation of the VUV light from its parent UV beams by suitable dichroics 
should effectively 
eliminate the high-recoil contribution, and would allow for the use of stronger VUV radiation to 
increase the ion yield while 
maintaining low overall recoil.

To illustrate the improvement in image resolution afforded by the implementation of the (1+1') 
REMPI 
scheme, the packet of C($^3P_1$) atoms exiting the decelerator with a mean longitudinal velocity of $\text{v}^\parallel=300$ m/s was 
velocity map imaged both with and without the addition of the $\lambda_{3,\text{res}}$ laser after VUV 
excitation, as well as with a conventional (2+1) REMPI detection scheme. The 
results are depicted in \autoref[(c - e)]{fig:REMPI}. For (2+1) REMPI detection, a 280.31 nm laser (10.5 mJ, 
partially 
focussed) was 
used to induce the 2p3p $^3P_1$ $\leftarrow$ 2p$^2$ $^3P_1$ transition and subsequently ionize the 
atom \cite{Bergstroem1989,Jankunas2015}, as schematically 
depicted in \autoref[(a)]{fig:REMPI}. The image noise 
arising from 
the ionization of background gas was found to be strongly increased for (2+1) REMPI detection in comparison with the VUV-based detection schemes. To suppress this noise, the $^{13}$C isotope was used when employing the (2+1) REMPI 
scheme, while the naturally most abundant $^{12}$C isotope was used for the images recorded 
with VUV radiation. Both isotopes are transmitted through the Zeeman decelerator with near-identical efficiency. The laser powers were attenuated for each beamspot image such that less than one ion per shot was recorded on average.
The 36 m/s recoil for the (2+1) REMPI detection of 
$^{13}$C causes the ion signal to appear on a ring centered around the 
mean velocity of the decelerated beam, and with a radius corresponding to the recoil velocity (see \autoref[(c)]{fig:REMPI}). The intensity distribution along the ring depends on the initial orbital of the ejected electrons as well as the laser polarizations. Similarly, when ionizing by $\lambda_1$ and $\lambda_2$ after VUV excitation, two concentric rings are 
observed 
that correspond to $^{12}$C recoil velocities of  
39 and 17 m/s, respectively (see 
\autoref[(d)]{fig:REMPI}). By contrast, when employing the $\lambda_{3,\text{res}}$ light for near-threshold ionization after VUV 
excitation, which reduces the $^{12}$C ion-recoil to $<0.9$ m/s, a small and well-defined spot is observed 
in the VMI image that reflects the velocity spreads of the Zeeman decelerated beam itself (see 
\autoref[(e)]{fig:REMPI}). The low VUV power generated by difference frequency mixing suppressed the absorption of another VUV photon 
after VUV excitation, and signals from this high-recoil ionization process could not be distinguished.

The near-threshold (1+1') REMPI scheme implemented here appeared remarkably efficient. The (1+1') scheme provided a similar ion yield as the conventional (2+1) REMPI scheme, while a strong decrease in ionization of background gas was observed. Together, this resulted in a significantly better signal to noise ratio in the scattering images captured with the (1+1') (VUV+UV) detection method. These observations show that, despite the low VUV power, VUV+UV detection as employed here can
provide a promising path to recoil-free detection of species for which near-threshold (UV+UV') REMPI schemes are either unavailable or experience strong (UV+UV) competition associated with large ion recoil.

\begin{figure}[!htb]
   \centering
    \resizebox{0.9\linewidth}{!}
    {\includegraphics{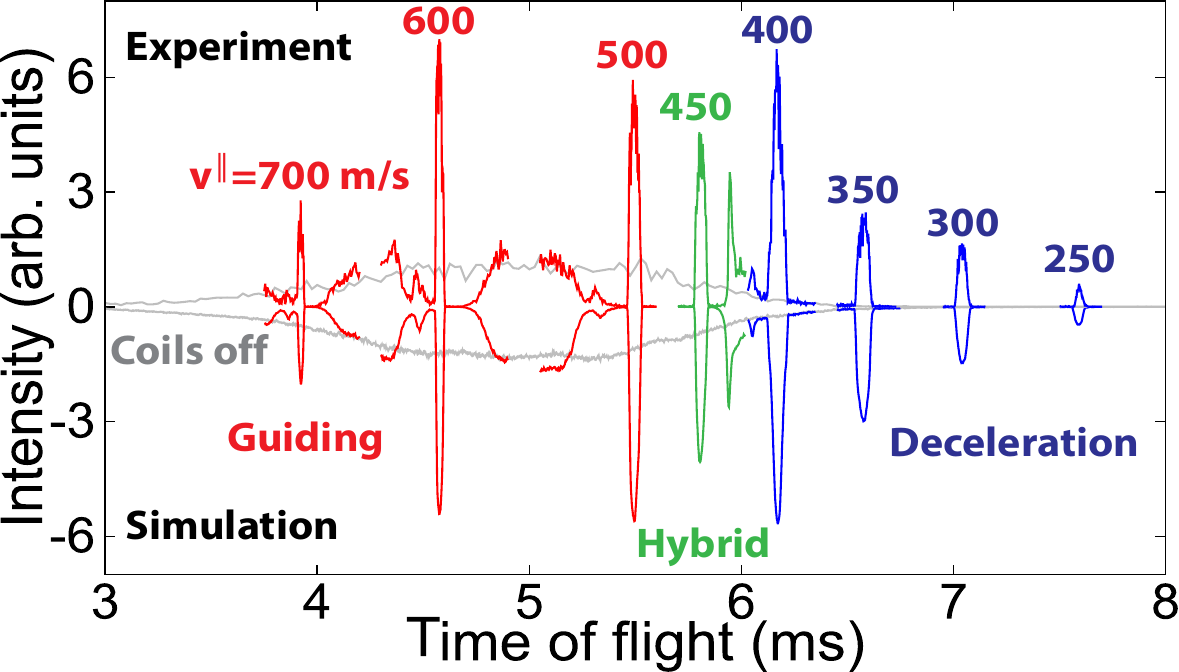}}
    \caption{Selected parts of the TOF profiles for C($^3P_1$) atoms that exit the decelerator when it 
    is programmed to either guide a packet of C($^3P_1$)-atoms at a constant speed (Guiding mode, red) or 
    decelerate a packet with an initial velocity of 500 m/s to various final velocities (Hybrid or 
    Deceleration modes, green and blue, respectively), see Experimental methods for further details. The experimental profiles are shown above the 
    simulated profiles, and the obtained longitudinal velocities ($\text{v}^\parallel$) corresponding 
    to the peaks are indicated. The profile recorded with the coils switched off is 
    shown for comparison (gray).}
    \label{fig:TOF}
\end{figure}

The packets of C($^3P_1$) atoms that exit the decelerator with various selected longitudinal 
velocities ($\text{v}^\parallel$) were characterized by recording their time-of-flight (TOF) 
profiles and by imaging their velocity distributions using VMI in combination with the low-recoil 
REMPI scheme. The TOF profiles are depicted in \autoref{fig:TOF} and show excellent agreement with the 
profiles obtained from numerical particle trajectory simulations that take into account the forces 
exerted on the C($^3P_1$) atoms by the space- and time-dependent fields inside our Zeeman decelerator 
apparatus. The VMI images were recorded at the peaks of the TOF profiles and thus capture the velocity 
distributions of the most intense part of the packets (see \autoref[(e)]{fig:REMPI} for the example of $\text{v}^\parallel=300$ m/s). From these images, the velocity distributions in 
both 
the longitudinal ($\parallel$) and transverse ($\perp$) direction were extracted and fitted with a 
Gaussian function. The resulting FWHM velocity 
spreads 
are summarized in 
\autoref{tab:spreads} and 
show good agreement with the values extracted from the simulations.

\begin{table}[!h] 
\centering
\caption{Experimental (Exp.) and simulated (Sim.) longitudinal (${\sigma_{\text{v}}}^\parallel$) and transverse (${\sigma_{\text{v}}}^\perp$) FWHM velocity spreads of the packets of C($^3P_1$) exiting the Zeeman decelerator with different mean longitudinal velocities ($\text{v}^\parallel$).  The possible contribution of residual ion-recoil to the experimental velocity spreads is not included in the simulations.}
\label {tab:spreads}
\begin{tabular}{lcccc}
\hline
 & \multicolumn{2}{c}{FWHM ${\sigma_{\text{v}}}^\parallel$} & \multicolumn{2}{c}{FWHM ${\sigma_{\text{v}}}^\perp$}\\
$\text{v}^\parallel$ (m/s) & \multicolumn{2}{c}{(m/s)} & \multicolumn{2}{c}{(m/s)}\\
{} & Exp. & Sim. & Exp. & Sim.\\
\hline
700 & 15.6 & 13.8 & 10.0 & 7.3\\
600 & 13.4 & 13.1 & 9.7 & 7.4\\
500 & 12.2 & 11.9 & 11.5 & 10.4\\
450 & 9.2 & 8.8 & 10.9 & 11.7\\
400 & 11.7 & 11.8 & 7.7 & 8.6\\
350 & 6.7 & 6.6 & 8.9 & 7.3\\
300 & 4.8 & 3.9 & 7.8 & 9.9\\
250 & 4.1 & 2.1 & 8.0 & 7.2\\
\hline
\end {tabular}
\end {table}

The well-controlled packets of Zeeman-decelerated C($^3P_1$) atoms in combination with the efficient 
low-recoil detection are an ideal starting point for a high-resolution crossed-beam scattering 
experiment. To demonstrate this, we recorded scattering 
images for de-excitation C($^3P_1$) + He $\rightarrow$ C($^3P_0$) + He collisions (see \autoref[(e)]{fig:C-He}) at three different 
collision energies ($E_{\text{coll}}$), see \autoref[(a - c)]{fig:C-He}.
The images are presented such that the relative 
velocity vector is directed horizontally, with forward scattering angles positioned at the right side 
of the image, see \autoref[(d)]{fig:C-He}. Small segments of the 
images are masked where the initial 
atomic beam gives a 
contribution to the signal. Besides the strong scattering ring corresponding to $^3P_1$ 
$\rightarrow$ $^3P_0$ de-excitation, a weak outer ring is observed for the 
$^3P_2$ $\rightarrow$ $^3P_0$ channel, which arises from the significantly lower 
density of C($^3P_2$) that is co-decelerated with the C($^3P_1$) atoms. The two rings are well 
separated due to the high image resolution.

\begin{figure*}[!htb]
   \centering
    \resizebox{1.0\linewidth}{!}
    {\includegraphics{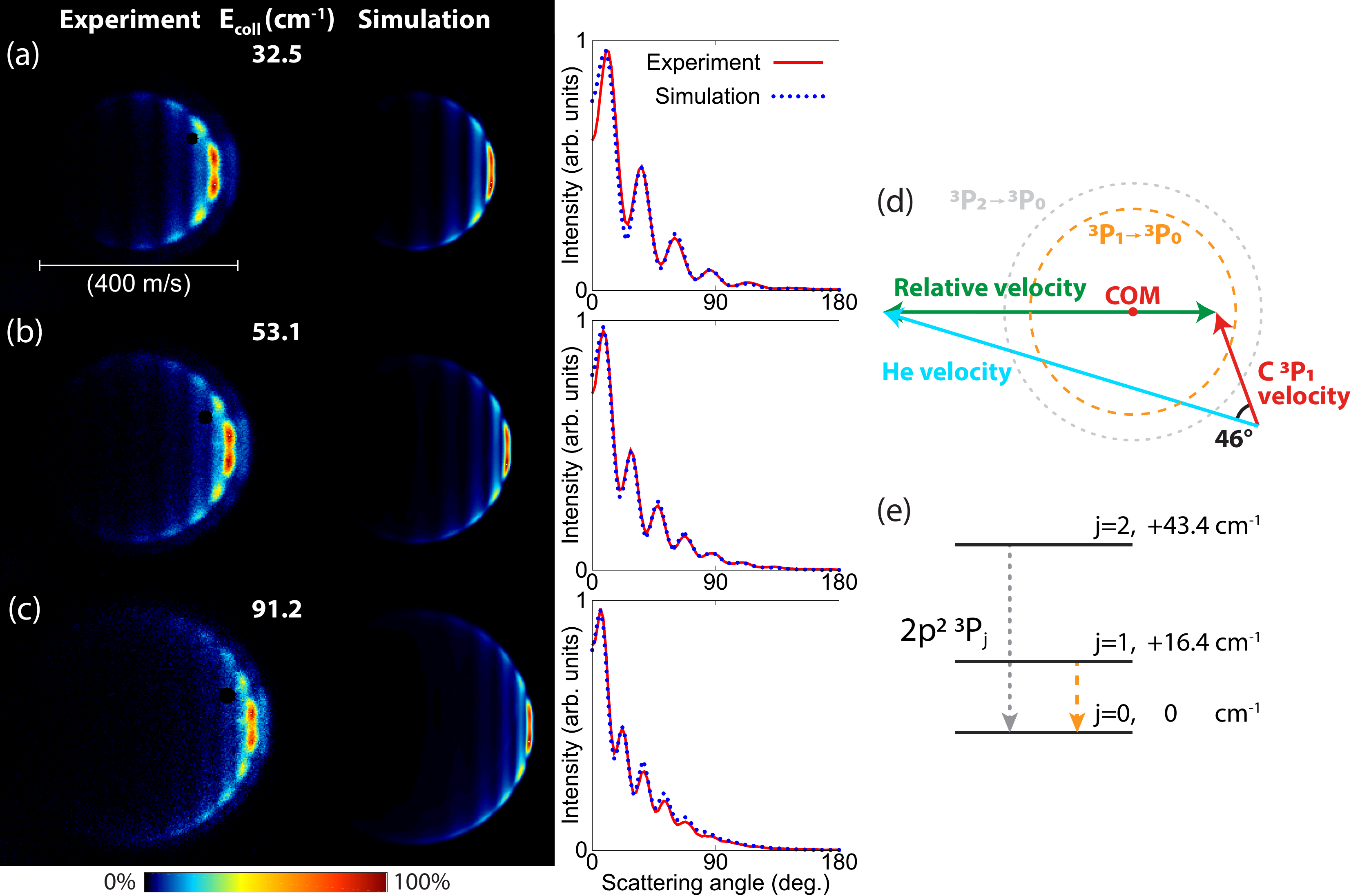}}
    \caption{(a)-(c) Experimental and simulated scattering images for the C($^3P_1$) + He $\rightarrow$ C($^3P_0$) + 
    He de-excitation process at three different collision energies, together with the corresponding angular scattering distributions extracted from the bottom half of each image. 
    One image pixel corresponds to a velocity of 2.41 m/s. The weak outer 
    rings  in the experimental images correspond to de-excitation from the C($^3P_2$) 
    initial state that is significantly less populated in the decelerated beam. Small segments of the experimental images are masked where the initial beam gives a contribution to the signal. (d) Schematic velocity 
    diagram illustrating how the scattered C atoms are projected on a circle around the center of mass 
    (COM). (e) Energy level diagram of the involved C($^3P_j$) spin-orbit levels.}
    \label{fig:C-He}
\end{figure*}

In each of the recorded scattering images, a clear oscillatory pattern can be observed by virtue of 
the exceptional resolution afforded by the combination of Zeeman-deceleration, VMI and near-threshold 
ionization used in the experiment \cite{VonZastrow2014}. These oscillations result from the quantum 
mechanical nature of the atoms that leads to 
diffraction of matter waves during the collision. The angular scattering distributions, shown in 
\autoref[(a-c)]{fig:C-He}, are retrieved from the experimental image intensities within a narrow annulus around the observed 
rings and can be directly compared to the distributions obtained from simulated images. Our image 
simulations are based on theoretical state-to-state cross sections acquired from 
quantum mechanical close-coupling (QM CC) calculations that use 
state-of-the-art \emph{ab initio} 
C-He 
PECs \cite{Bergeat2018,Bergeat2019,Klos2018}, in combination with the particle trajectory 
simulations on 
our Zeeman decelerator apparatus. The simulated images are shown next to the experimental ones, and are analyzed analogously to their experimental 
counterparts to acquire predicted angular scattering distributions that take into account the 
spatial, temporal and velocity spreads of the used atomic beams \cite{VonZastrow2014,Plomp2020}. 
Our measurements are in excellent agreement with the simulated distributions, which confirms the high 
quality of the PECs used in the scattering calculations. 

A qualitative understanding of the diffraction oscillations follows 
from a semi-classical picture in which a matter wave scatters on a 
structureless target. Within a hard-sphere model, the angular spacing between diffraction peaks can be 
approximated by $\Delta\theta=\pi/(kR)$, in which $k$ denotes the wavenumber of the incoming wave 
and $R$ is the radius of 
the sphere \cite{Onvlee2015}. The collision energy is related to $k$ 
through $\hbar k = \sqrt{2\mu E_\text{coll}}$, where $\mu$ is the reduced mass of the system. The sphere radius can be determined from the PECs as the C-He distance where the potential energy equals the experimental collision energy. The interaction of C($^3 P$) with He($^1 S$) gives rise to a doubly degenerate PEC of $^3\Pi$  character and a PEC of
$^3\Sigma$ character, which are coupled through the spin-orbit interaction \cite{Bergeat2018}. The values for $R$ and 
the values for $\Delta\theta$ predicted by a hard-shell model using both PECs are given in Table \ref{tab:diffosc} for the collision energies of the 
experiment, together with the values for $\Delta\theta$ that follow from the QM 
CC 
calculations. Qualitative agreement is found, both with respect to the values for $\Delta\theta$ and 
with respect to the observed trend that $\Delta\theta$ decreases with increasing collision energy.  

\begin{table}[!h] 
	\centering
	\caption{Parameters used for and following from the hard-shell model. For the experimental 
	collision energies ($E_\text{coll}$), the radius ($R$) of the sphere and the angular 
	spacing ($\Delta 
	\theta$) between the diffraction oscillations following from the QM CC 
	calculations and the hard-shell model with 
	the two potential energy curves (PECs) are listed.}
	\label {tab:diffosc}
	\small
\begin{tabular}{l c c c}
   \hline
   $E_\text{coll}$ (cm$^{-1}$) & 32.5 & 53.1 & 91.2 \\
   \hline
   $R$, $^3\Sigma$ PEC ($a_0$) & 5.0 & 4.8 & 4.7 \\
   $R$, $^3\Pi$ PEC ($a_0$) & 6.1 & 5.9 & 5.6 \\
   $\Delta \theta$ hard shell, $^3\Sigma$ PEC ($^\circ$) & 28 & 23 & 18 \\
   $\Delta \theta$ hard shell, $^3\Pi$ PEC ($^\circ$) & 23 & 19 & 15 \\          
   $\Delta \theta$ QM CC calculations ($^\circ$) & 26 & 20 & 15 \\   
   \hline
\end{tabular}
\end{table}

The ability to experimentally resolve detailed structures like diffraction oscillations in the angular scattering 
distributions of C + He collisions as demonstrated here shows that the combination of Zeeman deceleration, VMI and
near-threshold VUV REMPI detection allows for high-resolution 
measurements that provide a sensitive test for theoretical models. The resolution attained here is 
similar to the resolution achieved in crossed beam experiments that use Stark 
decelerated NO 
radicals, which currently defines the state of the art in this type of 
experiments 
\cite{VonZastrow2014, vogels2015imaging, gao2018observation, Jongh2020}. 

Our approach opens new vistas to study interesting collision phenomena in a wide variety of systems, for example, the observation of 
predicted scattering resonances in the ICSs 
and accompanying rapid changes of structure in the DCSs of low energy C($^3P_1$) + He 
\cite{Bergeat2018} as well as C($^3P_1$) + H$_2$ de-excitation collisions \cite{Bergeat2019,Klos2018}. 
Moreover, inelastic scattering of C($^3P$) 
atoms with complex molecules like O$_2$ and NO could be 
investigated to study, for instance, the role of nonadiabatic dynamics when open shell species interact, thus providing a further challenge for theory. Additionally, since a large variety of chemically relevant species is amenable to Zeeman or Stark deceleration \cite{Meerakker2012} and the use of VUV 
light provides the perspective of recoil-free REMPI detection for many 
species, the combination of techniques demonstrated here offers new and exciting prospects to study a 
large range of collision processes with an unprecedented level of precision. 
Noteworthy species like OH, CO, NH$_3$, and CH$_3$ possess well-known VUV 
transitions \cite{Beames2011,Sun2017,Perez2014,Teh2002}, although for some the intermediate state is strongly predissociative, and it remains to be seen how efficient and state-selective near-threshold multi-color REMPI schemes are best implemented for these species. 
Furthermore, making use 
of the recently reported near-threshold VUV REMPI schemes for H/D \cite{Yuan2018} or O($^3P$) atoms 
\cite{Wang2021}, our approach provides a pathway to high-resolution and low-energy investigations 
of elementary reactive scattering processes that produce O- or H-atoms, 
such as C + O${_2}$ $\rightarrow$ CO + O
\cite{Geppert2000,Karpov2020,Veliz2021} 
or complex-forming reactions between Zeeman decelerated atoms and H$_2$ 
molecules \cite{Aoiz2006, Guo2012, Yang2007}.

\section*{Experimental methods}
\begin{figure}[!htb]
   \centering
    \resizebox{1.0\linewidth}{!}
    {\includegraphics{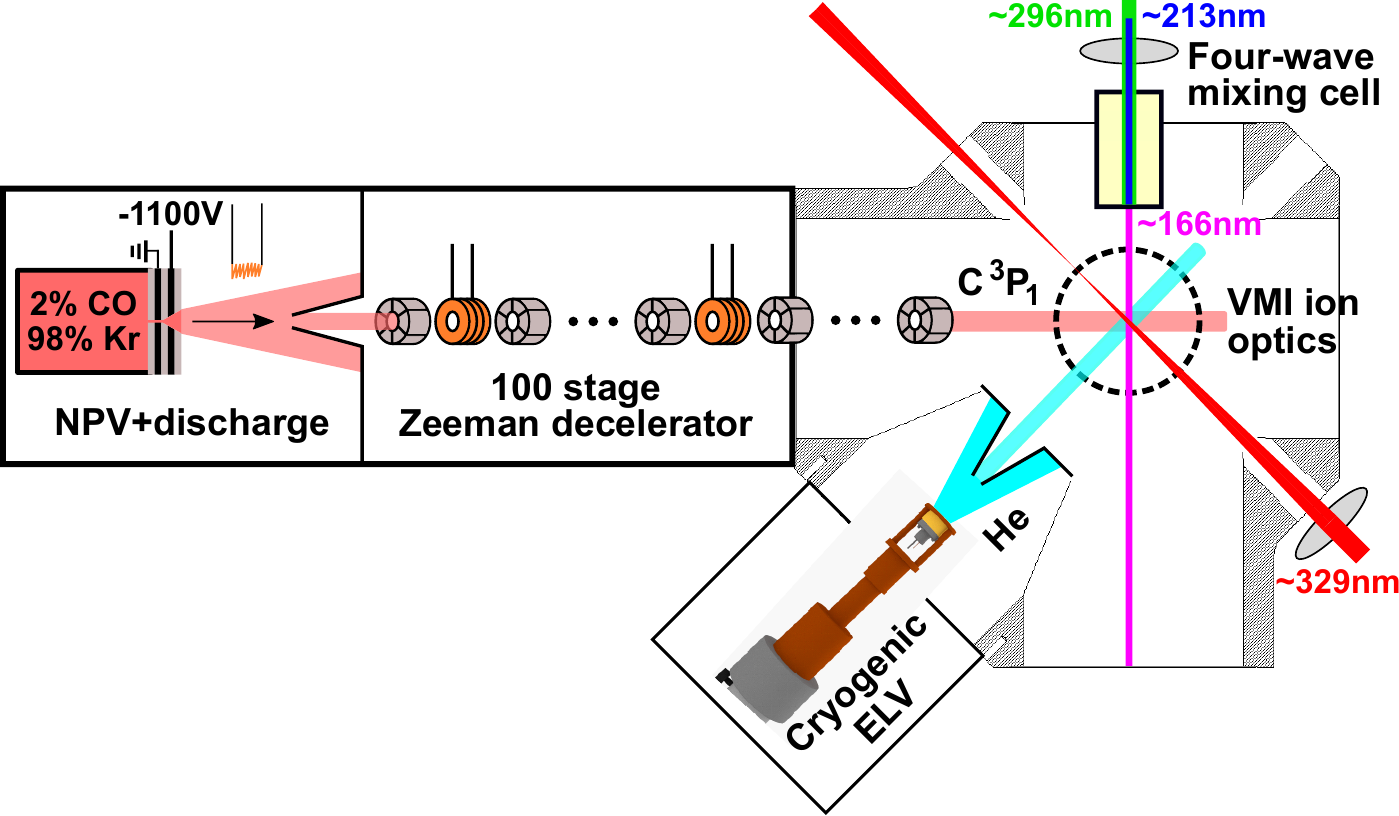}}
    \caption{Schematic depiction of the crossed-beam setup. The used combination of Zeeman 
    deceleration, (1+1') (VUV+UV) near-threshold REMPI detection and VMI allowed for high-resolution 
    imaging of 
    C-atom 
    scattering distributions after interaction with He.}
    \label{fig:Setup}
\end{figure}

A beam of carbon atoms, C($^3P_j$), with a mean velocity of around 
550 m/s was generated by running an 
electric discharge through an expansion of 2\% CO seeded in krypton (see \autoref{fig:Setup}), using a 
Nijmegen Pulsed Valve (NPV) with discharge assembly \cite{Ploenes2016}. After the expansion, the 
majority of the carbon atoms resided in the $^3P_0$ ground state spin-orbit level, while the 
$^3P_{1,2}$ levels were much less populated. This beam of carbon atoms then passed a skimmer and 
entered 
the Zeeman decelerator, of which a detailed description is given elsewhere \cite{Cremers2019}. 
Briefly, it consists of an alternating array of pulsed solenoids and permanent magnetic hexapoles that 
allow independent control over the longitudinal and transverse motion of paramagnetic species, 
respectively. The decelerator contains a total of 100 solenoids and 101 hexapoles, and was operated at 
a repetition rate of 20 Hz. Each coil can be pulsed once to either accelerate or decelerate the packet 
of C atoms as it passes the coil (Acceleration or Deceleration mode). Double pulses can be used 
(Hybrid mode), for example to increase contrast in the TOF profiles for mild deceleration or to guide 
the packet through the decelerator at a constant speed (Guiding) \cite{Cremers2019}. The C-atom 
$^3P_1$ state has a magnetic moment of 1.5 $\mu_B$ 
and splits into $m_j=0, \pm1$ components in the presence of a magnetic field, with $m_j$ the 
projection of the total electronic angular momentum $j$ on the space-fixed $z$-axis. The 
$m_j$ = 1 component was effectively manipulated with the decelerator. Similarly, the $^3P_2$ 
state 
has a magnetic moment of 3 $\mu_B$ and splits into five components, i.e., $m_j=0, \pm 1, 
\pm 
2$. 
Although the C($^3P_2, m_j=1,2$) components were co-decelerated with the C($^3P_1, m_j=1$) atoms, 
their 
density in the beam was significantly lower. While the $^3P_0$ state had a much higher initial 
population, it only has an $m_j=0$ component which is almost insensitive to magnetic fields. The 
resulting free flight through the decelerator heavily reduced the $^3P_0$ atom density, such that its 
final contribution was negligible. Thus, after the decelerator a beam 
of mainly C($^3P_1$) atoms was obtained with controlled velocity and narrow angular and velocity 
spreads. A series of 13 additional hexapoles guided the packets of C($^3P_1$) atoms towards the interaction 
region, where they were intersected by a beam of He atoms at an angle of 
$46^{\circ}$ about 368.5 mm from the decelerator exit. The He beam was produced using an Even-Lavie 
Valve (ELV) that was cryogenically cooled to control the mean velocity, thus changing the mean 
collision energy to 32.5, 53.1, or 91.2 cm$^{-1}$ when intersected by the packets of C($^3P_1$) atoms 
that were decelerated to a final velocity of 350 m/s. After scattering, the product C($^3P_0$) 
atoms were state-selectively ionized using a near-threshold (1+1') (VUV+UV) REMPI scheme, and detected with the use of 
high-resolution VMI ion optics that allows for accurate mapping of large ionization volumes \cite{Plomp2021}. Due to the obtained narrow velocity spreads of the decelerated C atoms the scattering signal arising from the contribution of co-decelerated initial C($^3P_2$) could be well separated from the main C($^3P_1$) contribution.

\section*{Theoretical methods}
In a full description of the collision process including the electronic fine-structure of the C atom, 
the 
states in the C($^3P_j$)+He arrangement are described by the quantum number $j$, which corresponds to 
the total 
electronic 
angular momentum of the $^3P$ carbon atom ($\bf{j} = \bf{L} + \bf{S}$ with $L$ and $S$ the 
electronic orbital 
and spin angular momenta, respectively). For the calculation of the integral and differential cross 
sections for the collision of C($^3P_j$) with He to give C($^3P_{j'}$), we used 
the close coupling 
approach of Pouilly \etal~\cite{Pouilly:85} implemented in the HIBRIDON package \cite{Hibridon}. The 
calculations were performed with C-He PECs of Bergeat \etal~\cite{Bergeat2018} 
calculated using the spin-unrestricted single and double excitation coupled cluster approach with 
non-iterative perturbational treatment of triple excitations (UCCSD(T)) \cite{knowles:93} and an
augmented correlation-consistent polarized valence quintuple-zeta (aug-cc-pV5Z) basis set completed 
with additional 3s 3p 2d 2f 1g mid-bond functions \cite{dunning:89}. The asymptotic experimental 
spin-orbit splitting of C($^3P$) ($A_{SO}=\Delta_{j=1}=16.41671$ cm$^{-1}$ and $\Delta_{j=2}=43.41350$ 
cm$^{-1}$) \cite{Harris2017} was used in the computation of energy levels and in the quantum scattering calculations. In 
all calculations, the propagation was 
performed for $R$ ranging from 2.5 to 80 Bohr, with $R$ the interatomic C-He distance. 
The reduced mass of the C-He complex is $\mu_{r} = 3.001$ u. At each collision 
energy, the maximum 
value 
of the total angular momentum $J_{\text{max}}$ was set large enough to converge the integral and 
differential 
cross sections within 0.001 \r{A}$^2$. The effective 
DCSs that were used as input for the image simulations were constructed from the computed DCSs by 
taking into account the experimental collision energy spreads as a Gaussian distribution.

\section*{Acknowledgments}
This work is part of the research program of
the Dutch Research Council (NWO). This project has received funding from the European Research Council (ERC) under the European 
Union's Horizon 
2020 research and innovation program (Grant Agreement No. 817947 FICOMOL). It has 
moreover received funding from the European Union's Horizon 2020 research and 
innovation program under the Marie Sk{\l}odowska-Curie grant agreements No. 886046 and 889328. We 
thank Xingan Wang for discussions and support on VUV generation, and David Parker for discussions on REMPI 
of carbon atoms and careful reading of the manuscript. We thank Niek Janssen, Andr\'e van Roij and Edwin Sweers for expert technical support. 

\bibliography{2021_C-He_Zeeman}

\end{document}